\documentclass[english]{hermes-journal}

\usepackage{tabu}
\usepackage{array,booktabs}
\usepackage[table]{xcolor}
\usepackage{hyperref}
\usepackage{natbib}
\usepackage{bm}

\rmpd{2014}{28}{2-3}

\firstpagenumber{1}

\title[Collaborative network for live show recommendation]{Propagation of content similarity through a collaborative network for live show recommendation}

\author[1]{Jean}{Creusefond}
\author[2]{Matthieu}{Latapy}

\address{Delight\\33 Rue LaFayette\\
  75009 Paris, France}
{jean@delight-data.fr}
\address{Sorbonne Universités, UPMC Univ Paris 06\\
  CNRS, LIP6 UMR 7606\\
  4 place Jussieu\\
  75005 Paris, France}
{Matthieu.Latapy@lip6.fr}

\abstract{
  
  We present a network-based recommender system for live shows (concerts, theater, circus, etc) that finds a set of people probably interested in a given, new show.
  We combine collaborative and content-based filtering to take benefit of past activity of users and of the features of the new show.
  Indeed, as this show is new we cannot rely on collaborative filtering only.

  To solve this cold-start problem, we perform network alignment and insert the new show in a way consistent with collaborative filtering.
  We refine the obtained similarities using spreading in the network.
  We illustrate the performances of our system on a large scale real-world dataset.
}


\rowcolors{2}{gray!25}{white}

\begin{document}

\maketitle

\newpage

Delight\,\footnote{\url{http://delight-data.com/}} aims at finding people that will probably be interested in a given (new) live show.
To do so, it relies on longitudinal recordings of tickets previously bought by clients, and on features describing (past and new) shows.
Delight therefore faces a recommendation task, which is classically addressed with two families of approaches.

On the one hand, \textbf{content-based filtering}~\citep{ricci_content-based_2011} focuses on similarity between show features: if the user bought tickets for shows with features similar to the ones of the new show then he/she will probably be interested in the new show.
This approach does not depend on behaviors of other clients, but it relies on accurate and comparable show descriptions, which are rarely available.
It is known to lead to recommendations that lack diversity~\citep{ricci_content-based_2011}.

On the other hand, \textbf{collaborative filtering}~\citep{sarwar_item-based_2001} relies on correlations between tickets bought by different users: if two users bought many tickets for the same shows, then a new show of interest for one of them probably is of interest for the other too.
Such systems may recommend items even though they are different in nature: \textit{e.g.} you purchased a camera, and then a camera cover is recommended because both items are often purchased by the same people.

Collaborative filtering is appealing, but it suffers from a serious limitation in our context: as we are searching for people interested in a {\em new} show.
Therefore, we cannot take advantage of correlations between buyers of tickets for this show and buyers of tickets for other shows.
This is a typical {\em cold-start} problem, meaning that the data needed to make recommendations is not yet available.

In this paper, we introduce a recommender system that relies on collaborative filtering but uses content-based filtering for solving our cold-start problem, thus having the advantages of both approaches.
We first present the system itself and we present experiments performed on Delight data.

\section{Our recommender system}



Our system relies on the following three steps : we first build a collaborative item-item network (a graph where nodes are items), we then insert the new show using its features, and finally we perform recommendation by similarity propagation.

\subsection{The item-item collaborative network}

Given two shows $s_1, s_2$ from a set $S$ of shows, we consider a co-purchase measure denoted by $w(s_1,s_2)$.
The role of this function is to quantify the similarity between shows $s_1$ and $s_2$ regarding the people that bought tickets for them.
Several functions capture this intuition differently.
Choosing the best function is non-trivial, we will instead consider several ones proposed in the literature and compare their performances.

We denote by $U(s)$ the set of users that bought item $s$ and $S(u)$ the set of shows bought by user $u$.
We also denote by $U(s_1, s_2) = U(s_1)\cap U(s_2)$, set of users that bought items $s_1$ {\em and} $s_2$.
Finally, we introduce $k_{u} = |S(u)|$ and $k_{s} = |U(s)|$ for all user $u$ and show $s$.
With these notations, we list the considered functions in Table~\ref{tab:weight-func}.

Notice that co-purchase measures may be symmetric ($\forall (s_1, s_2)$, $w(s_1,s_2) = w(s_2,s_1)$) or asymmetric.
In our context, it may be important to capture asymmetric relations, as purchasing a ticket for show $s_1$ may give information on ones interest for show $s_2$ that is different than the converse.
We therefore considered asymmetric variants of the symmetric similarity functions.

\noindent
\tabulinesep=1.5mm
\begin{table}[tab:weight-func]{The tested weight functions}
  \centering
  \begin{tabu} {>{\bfseries}c|c}
    Amazon~\citep{smith_two_2017} & $\frac{|U(s_1, s_2)| - \sum_{u\in U(s_1)} 1 - (1 - k_{s_2}/\sum_{s\in S}k_s)^{|S(u) \backslash s_1|}}{\sqrt{|U(s_1, s_2)|}}$\\
    Jaccard & $\frac{|U(s_1, s_2)|}{|U(s_1) \cup U(s_2)|}$\\
    Jaccard-asym & $\frac{|U(s_1, s_2)|}{k_{s_1}}$ \\
    NBI~\citep{fiasconaro_hybrid_2015} & $\frac{1}{k_{s_1}} \sum_{u\in U(s_1, s_2)}\frac{1}{k_u}$ \\
    MDW~\citep{fiasconaro_hybrid_2015} & $\frac{1}{max(k_{s_1}, k_{s_2})}\sum_{u\in U(s_1, s_2)}\frac{1}{k_u - 1}$ \\
    MDW-asym & $\frac{1}{k_{s_1}}\sum_{u\in U(s_1, s_2)}\frac{1}{k_u - 1}$ \\
    BP~\citep{fiasconaro_hybrid_2015} & $\frac{|U(s_1, s_2)| - k_{s_1}  k_{s_2} / |S|}{\sqrt{k_{s_1}  (1 - k_{s_1} / |S|)\times  k_{s_2}  (1 - k_{s_2} / |S|)}}$ \\
  \end{tabu}
\end{table}

Most pairs of shows have no client in common, and so in most cases $w(s_1,s_2) \le 0$.
It is therefore most natural to see the similarity as a weighted network in which nodes are the shows, links are the pairs of shows with positive similarity, and the weight function is $w$.
We call this network the {\em collaborative item-item network}.

\subsection{Predicting weight from content}
\label{subsubsec:learn}

Due to the cold-start problem, we then rely on a content-based approach to insert a new show in the network.
We compare the features of the new show and the features of previous shows, and we link the new show to its more similar ones.

We need this insertion to be consistent with the collaborative network; more precisely, we expect the links of the new show and their weight to be similar to the ones we would obtain with a collaborative approach.
To ensure this, we learn a relationship between content-based weights and the collaborative ones.
Indeed, we have both values for already known shows.
By assuming a correlation between them, we can learn parameters in order to predict a collaborative similarity from the content-based one.

In practice, we follow the strategy proposed in \citet{debnath_feature_2008}:
\begin{enumerate}
\item create a set $E$ of edges from positive edges $w(s_1, s_2) > 0$ and from a sample of negative edges $w(s_1, s_2) \le 0$. On the one hand, including all negative edges would be computationally expensive, but on the other hand they are needed to learn what negative events are. We used a sample of the same size as the set of positive edges.
\item \label{item:cont-sim}compute a content similarity between features (\textit{e.g.} for cities, $1$ if the cities are the same and $0$ otherwise). Create a matrix $A$ where each line is an edge and each column is a feature category. Values are content similarities.
\item create a predictive model from the training matrix $A$ and the weight of the corresponding edges as a target variable. We used a linear regression model computed with a stochastic gradient descent algorithm. 
\end{enumerate}

When inserting a new show $s_{new}$, we use this predictive model for computing the weight of the edges between $s_{new}$ and all the other shows.

\subsection{Similarity propagation}

We then propagate the collaborative similarities in order to explore the neighborhood of the new show.
Let us call $t_l(s)$ the item-item similarity between $s_{new}$ and any show $s$ at the step $l$ of the propagation.

We start with only the new show being similar to itself $t_0(s_{new}) = 1$ and $\forall s\in S, t_0(s) = 0$.
Then, we update the similarities using the following formula :
\begin{equation}
  \label{eq:prop}
  t_{i+1}(s) = \frac{\sum_{s'\in S} t_i(s') \times w'(s', s)}{\sum_{s'\in S} t_{i+1}(s')}
\end{equation}

This propagation is designed so that all similarities at a given step sum to one.
Without propagation ($l = 1$), it is equivalent to \citet{debnath_feature_2008}.


Once the item-item similarity is computed for the new show, we can finally compute the preference function for each user towards the new show :
\begin{equation}
  r(u, s_{new}) = \max_{s\in S(u)} \sum_{1 \le i \le l} t_{i}(s)
\end{equation}

We sum the values of the different steps of the propagation in order to mitigate alternating scores (\textit{e.g.} if $s_{new}$ is connected to only one node $s$, we would have $t_l(s) = 0$ for odd $l$ and $t_l(s) = 1$ otherwise).
The preference of the user is defined as the maximum of the summed similarities of the shows that he saw in order to favor users that saw shows that are very similar to $s_{new}$.
Other aggregations (such as a sum) have the tendency of favoring users that saw many shows that may be of low relevance.

\section{Experiments}

This section present practical experiments that illustrate the impact of similarity propagation and the impact of the chosen weight function.

\subsection{Experimental protocol}

Our dataset is composed of 4.2 millions users that bought one or several tickets in 8 millions transactions for 120 thousands shows.
In addition, we have show descriptions obtained from several ticket sellers websites and APIs.
Using feature selection, we narrowed these descriptions to four key features: city, venue, types (rock, concert, ...) and stakeholders (artists, directors, ...)

We evaluate that an audience is relevant for a user of our system if that audience would spend a lot on the new show.
However, reaching people is costly.
We estimated that, given the state of business, the communication cost is $0.05$ euros.
We consider user revenue as a predictability indicator of an audience (amount payed by spectators for this show minus the communication cost).
Therefore, we evaluate the recommender system results by the optimal revenue audience that it could output.

We use the traditional holdout method as an evaluation strategy.
We first separate the set of shows into an historical training set and a more recent testing set.
Then, we train each version of our algorithm on the training set and run predictions for 96 shows of the testing set.
The produced rankings are then evaluated by optimal revenue that is averaged over the 96 test shows.

\subsection{Results}


\begin{table}[tab:grid-search]{Grid search for maximum revenue (in euros) with propagation length and weight function as parameters}
  \centering
  \begin{tabular}{r|ccccccc}
    \rowcolor{gray!50}
    & Amazon & BP & Jaccard & Jaccard-asym & MDW & MDW-asym & NBI \\
    $l=1$ & 843 & 848 & 850 & 890 & 856 & 913 & 913 \\
    $2$ & 454 & 794 & 773 & 881 & 876 & \textbf{923} & 913 \\
    $3$ & 411 & 770 & 737 & 800 & 824 & 900 & 867 \\
    $4$ & 326 & 765 & 715 & 779 & 817 & 840 & 804 \\
    $5$ & 282 & 703 & 671 & 756 & 798 & 812 & 799 \\
  \end{tabular}
\end{table}

We first note from Tab.~\ref{tab:grid-search} that using similarity propagation for our problem is beneficial to some extent.
Indeed, the top revenue with $l>1$ ($923$) is slightly above the top revenue with $l=1$ ($913$).

Secondly, we observe that there is a clearer benefit of comparing different weight functions ($843$ for the worst function, $923$ for the best).
We also note that using asymmetric versions of weight functions (Jaccard and MDW) yields better revenue that using their symmetric counterpart.
We think that an asymmetric similarity may reflect the fact that attending niche shows is more distinctive than attending popular ones.

\section{Related work and discussion}

The design of hybrid recommender systems has been studied by \citet{burke_hybrid_2007}.
In his paper, he describes different strategies for combining recommender systems of different natures.
While close in spirit, our algorithm is not a direct application of these methodologies.
Indeed, \citet{burke_hybrid_2007} uses the algorithms as black boxes, while we combine concepts that are internal to these algorithms.

The diffusion of similarities through an item-item network is the basis of the \textit{ItemRank} algorithm~\citep{gori_itemrank_2007}.
They propagate similarities by using a modified PageRank where items rated by the user are considered as points with fixed teleportation probabilities.

Feature weighting has been used by \citet{debnath_feature_2008} to create a hybrid algorithm.
They use machine learning to predict an item-item collaborative similarity from content-based similarities.
We push their work further by experimenting with multiple similarity scores and propagating them through an item-item network.

This paper presents development on a particular problem : ranking users for a specific product.
However, our algorithm could be transposed to the task of ranking items for a specific user.
One could imagine introducing new shows to an item-item network and start a similarity propagation from the shows seen by the user.

We also note that our learning strategy cannot capture complex interactions between instances of features.
For instance, the system cannot learn as-is that two shows might share an audience because two specific cities are close to each other.


\bibliography{biblio}

\end{document}